\documentclass[twolecolumn]{elsart3p}
\newcommand{\vecbm}[1]{\mbox{\boldmath $#1$}}

\usepackage{amssymb}
\usepackage[dvips]{graphicx}
\usepackage{bm}% bold math
\usepackage{amsmath}
\journal{Physica C}

\begin{document}
%\noindent
%PC-13 / ISS2014

%\noindent
%Submitted 31 January 2015\\
%Submitted 2 February 2015\\
\begin{frontmatter}


%5 \author{Name\corref{cor1}\fnref{label2}}

\title{Excitation spectra and wave functions of quasiparticle bound states in bilayer Rashba superconductors}

\author[AA]{Yoichi Higashi\corauthref{cor1}},\,
\ead{higashiyoichi@ms.osakafu-u.ac.jp}\,
\author[BB]{Yuki Nagai},\,
\author[CC]{Tomohiro Yoshida},\,
\author[AA]{Masaru Kato},\,
\author[DD]{Youichi Yanase}

\address[AA]{Department of Mathematical Sciences, Osaka Prefecture University, 1-1 Gakuen-cho, Naka-ku, Sakai 599-8531, Japan}
\address[BB]{CCSE, Japan Atomic Energy Agency, 178-4-4, Wakashiba, Kashiwa, Chiba 277-0871, Japan}
\address[CC]{Graduate School of Science and Technology, Niigata University, Niigata 950-2181, Japan}
\address[DD]{Department of Physics, Niigata University, Niigata 950-2181, Japan}

\corauth[cor1]{Corresponding author.
Department of Mathematical Sciences, Osaka Prefecture University, B9 Bldg., 1-1 Gakuen-cho, Naka-ku, Sakai 599-8531, Japan
Tel.: +81-72-254-9368 ; fax: +81-72-254-9916}

\begin{abstract}
We study the excitation spectra and the wave functions of quasiparticle bound states at a vortex and an edge in bilayer Rashba superconductors under a magnetic field.
In particular, we focus on the quasiparticle states at the zero energy in the pair-density wave state in a topologically non-trivial phase.
We numerically demonstrate that the quasiparticle wave functions with zero energy are localized at both the edge and the vortex core
if the magnetic field exceeds the critical value.

%It is the purpose of this study to clarify the energy dependence of the local density of states (LDOS) around a multi-quantum vortex in the presence of non-magnetic impurities. 
%We calculate the energy and the spatial dependence of the LDOS in the vicinity of an even winding-number vortex for several values of the impurity scattering rate.
\end{abstract}

\end{frontmatter}
% \linenumbers

\section{Introduction}
%\label{}

Superconductivity in locally noncentrosymmetric systems has recently attracted considerable interest \cite{Yoshida2014,Sigrist2014}.
In such systems, a variety of anti-symmetric spin-orbit coupling arises depending on the local inversion symmetry breaking.
A lot of materials have the local noncentrosymmetricity in their crystal structures,
one of which is the artificially fabricated heavy Fermion superlattice CeCoIn$_5$/YbCoIn$_5$ \cite{Mizukami2011}.
Some experiments have been conducted under a magnetic field, focusing on the role of the noncentrosymmetric superconductivity in the superlattice CeCoIn$_5$/YbCoIn$_5$ \cite{Goh2012,Shimozawa2014}.
From the experiment on the angular dependence of the $H_{\rm c2}$,
Goh $et$ $al$. obtained the evidence that the FFLO like inhomogeneous superconductivity realizes
under the parallel field in the CeCoIn$_5$/YbCoIn$_5$ superlattice \cite{Goh2012}.
The superconductivity in multilayered systems as a simple model of the superlattice of CeCoIn$_5$ is investigated also theoretically \cite{Maruyama2012}
and the spatially inhomogeneous exotic superconducting phase is proposed under a high magnetic field \cite{Yoshida2012}.

In our previous study,
we analyzed the electronic structure of a single vortex in the pair-density wave (PDW) state,
in which the order parameter phase changes its sign in bilayer systems.
Then we found the salient feature of the PDW state, that is,
the zero energy quasiparticle states at the vortex core exist even in a high magnetic field,
using the quasiclassical theory \cite{Higashi2014}.
However, we could not obtain the more microscopic information on the vortex core structure
such as the excitation spectra and the wave functions, in principle,
in the framework of the quasiclassical theory.

In this study, we do not touch the BCS phase, in which the order parameter phase does not change its sign,
and focus on the PDW state.
We obtain the microscopic information on the quasiparticle states in the PDW state
with use of the Bogoliubov-de Gennes theory.
We present the excitation spectra and the Bogoliubov quasiparticle wave functions
of bound states at the vortex core and the edge.
%\\
%* in the absebce of the SOC, this zero energy states is removed by the magnetic field (Zeeman split)\\
%* in the presence of the sufficiently large SOC, the zero energy state is not removed by the magnetic field. \\
%%%%%%%%%%%%%%%%%%%%%%%%%%%%%%%%%%%%%%%%%%%%%%%%%%%%%%%%%%%%%%%%%%%%%%%%%%%%%%%%%%%%%%%%%%%%%%%%%%%%%%%%%%%%%%%%%%%%%%%%%%%%%%%%%%%%%%
%%%%%%%%%%%%%%%%%%%%%%%%%%%%%%%%%%%%%%%%%%%%%%%%%%%%%%%%%%%%%%%%%%%%%%%%%%%%%%%%%%%%%%%%%%%%%%%%%%%%%%%%%%%%%%%%%%%%%%%%%%%%%%%%%%%%%%
\section{Formulation}
First of all, we characterize multilayer Rashba systems under a magnetic field.
We incorporate the effect of the magnetic field into the theory
through the Zeeman term reflecting the dominant paramagnetic effect
and the spatial inhomogeneity of the order parameter due to a vortex line.
Reflecting the lacking of mirror symmetry about each layer,
the antisymmetric spin-orbit coupling (ASOC) strength has the layer dependence $\alpha_m$ with the layer index $m$.
The layer dependence of the ASOC is $(\alpha_1,\alpha_2)=(\alpha,-\alpha)$ in the bilayer system.
The net ASOC is zero due to the mirror symmetry of the entire system.

We start with the following Bogoliubov-de Gennes equation for bilayer Rashba superconductors.
%%%%%%%%%%%%%%%%%%%%%%%%%%%%%%%%%
\begin{align}
\left(
\begin{array}{cc}
\hat{H}\left( -i{\vecbm \nabla} \right) & \hat{\varDelta}({\vecbm r}) \\
\hat{\varDelta}^\dag({\vecbm r}) & -\hat{H}^\ast\left( i{\vecbm \nabla}) \right)
\end{array}
\right)
&
\left(
\begin{array}{c}
{\vecbm u}_{jn}({\vecbm r})  \\
{\vecbm v}_{jn}({\vecbm r})
\end{array}
\right)
\nonumber \\
=
E_{jn}
&
\left(
\begin{array}{c}
{\vecbm u}_{jn}({\vecbm r})  \\
{\vecbm v}_{jn}({\vecbm r})
\end{array}
\right),
\label{BdG}
\end{align}
%%%%%%%%%%%%%%%%%%%%%%%%%%%%%%%%%
with the normal state Hamiltonian in the real space representation
%%%%%%%%%%%%%%%%%%%%%%%%%%%%%%%%%
\begin{align}
\hat{H}\left( -i{\vecbm \nabla} \right)&=
\left(
\begin{array}{cc}
h_1(-i{\vecbm \nabla}) & t_\perp \sigma_0  \\
t_\perp \sigma_0 & h_2(-i{\vecbm \nabla})
\end{array}
\right),\\
%%%%%%%
h_1(-i{\vecbm \nabla})&=\xi \left( -i{\vecbm \nabla}  \right)\sigma_0 -\mu_{\rm B} {\vecbm H} \cdot {\vecbm \sigma} +\alpha_1 {\vecbm g} \left(-i{\vecbm \nabla}\right)\cdot {\vecbm \sigma},\\
h_2(-i{\vecbm \nabla})&=\xi \left( -i{\vecbm \nabla}  \right)\sigma_0 -\mu_{\rm B}{\vecbm H} \cdot {\vecbm \sigma} +\alpha_2 {\vecbm g} \left( -i{\vecbm \nabla} \right)\cdot {\vecbm \sigma},
\end{align}
%%%%%%%%%%%%%%%%%%%%%%%%%%%%%%%%%
where ${\vecbm u}_{jn}({\vecbm r})$ and ${\vecbm v}_{jn}({\vecbm r})$ are the wave functions of Bogoliubov quasiparticles
with $n$ and $j$ the angular momentum (azimuthal) and the radial quantum number, respectively,
and $\xi(-i{\vecbm \nabla})=(-i{\vecbm \nabla})^2/(2m)-\mu$ with the mass of electron $m$ and the chemical potential $\mu$.
$h_1$ and $h_2$ are the normal state Hamiltonian for the layer 1 and 2, respectively.
$\mu_{\rm B}$ is the Bohr magneton, ${\vecbm H}$ is the magnetic field vector, ${\vecbm \sigma}=(\sigma_x,\sigma_y,\sigma_z)^{\rm T}$ is the vector of the Pauli spin matrices, $\sigma_0$ is the unit matrix in the spin space
and $t_\perp$ is the interlayer hopping strength.
We take the Rasba type ASOC characterized by the orbital vector ${\vecbm g}(-i{\vecbm \nabla})=(i\partial_y,-i\partial_x,0)/k_{\rm F}$ with the Fermi wave number $k_{\rm F}$.
We use the unit in which $\hbar=1$.

We consider a single vortex in a spin-singlet $s$-wave disk-shaped superconductor with its radius $r_{\rm c}$.
The magnetic field is applied perpendicular to the layer ${\vecbm H}=(0,0,H)$ and the vortex line is parallel to the $z$ axis.
The pair potential in the bilayer system is expressed as
%%%%%%%%%%%%%%%%%%%%%%%%%%%%%%%%%
$
\hat{\varDelta}({\vecbm r})=\varDelta_0 f(r) e^{i\phi_{\rm r}}
{\rm diag}(i\sigma_y, si\sigma_y)
$
%%%%%%%%%%%%%%%%%%%%%%%%%%%%%%%%%
where $s=1$ $(s=-1)$ denotes the BCS (PDW) state.
The vortex center is situated at the origin ${\vecbm r}={\vecbm 0}$.
We assume the spatial profile of the pair potential as the same in each layer
and put $f(r)=\tanh(r/a)$%$f(r)=\tanh(r/\xi_0)$
around a vortex.
%We define the coherence length $\xi_0=v_{\rm F}/\Delta_0$ with the Fermi velocity $v_{\rm F}$.
Here, $a$ is the lattice constant, which is introduced as follows.
Expanding the tight-binding model for the two dimensional square lattice, $\varepsilon(\bm{k})=-2t[\cos(k_x a)+ \cos(k_y a)]$
with respect to $k_x$ and $k_y$ up to the power of two
and comparing with the free electron dispersion $\varepsilon(k)=k^2/2m$,
we have the relation, $t=1/(2ma^2)$.
We have introduced the lattice constant $a$ as a unit of length and the nearest neighbor hopping integral $t$ as a unit of energy.
The characteristic length scale of superconductivity is the coherence length,
but in the present study, a Fermi surface is split due to the SOC, the Zeeman field and the inter layer hopping,
and then the coherence length depends on them.
To avoid this difficulty,
we use $a$ and $t$ as the unit of length and energy, respectively.
They are independent of the SOC, the Zeeman field and the inter layer hopping.
%We define the lattice constant $\xi_0=v_{\rm F}/\Delta_0$ with the Fermi velocity $v_{\rm F}$.

The system has the rotational symmetry about a vortex line.
So we can introduce the cylindrical coordinates.
In this situation, we can separate the quasiparticle wave functions into the angular and the radial parts:
${\vecbm u}^i_{jn_i}({\vecbm r})=\exp(i n_i \phi_{\rm r}){\vecbm u}^i_j(r)$,
${\vecbm v}^i_{jl_i}({\vecbm r})=\exp(i l_i \phi_{\rm r}){\vecbm v}^i_j(r)$ ($1 \leq i \leq 4$, $i=\{\sigma,m\}$) with the spin index $\sigma$.
Substituting these wave functions into the BdG equation (\ref{BdG}) in the cylindrical coordinates,
one can find the relation with respect to the quantum number of the orbital angular momentum such that
$n_1=n_3=l_1=l_3\equiv n$, $n_2=n_4=n+1$, $l_2=l_4=n-1$.
Then, we obtain the following dimensionless BdG equation.
%%%%%%%%%%%%%%%%%%%%%%%%%%%%%%%%%
\begin{align}
&\left(
\begin{array}{cc}
\hat{H}_n(r,\partial_r) & \hat{\Delta}(r) \\
-\hat{\Delta}(r) & -\hat{H}_{-n}(r,\partial_r)
\end{array}
\right)
\left(
\begin{array}{c}
{\vecbm u}_j(r)  \\
{\vecbm v}_j(r)
\end{array}
\right)\nonumber \\
&~~~~~~~~~~~~~~~~~~~~~~~~~~~~~~
=
E_{jn}
\left(
\begin{array}{c}
{\vecbm u}_j(r)  \\
{\vecbm v}_j(r)
\end{array}
\right),
\end{align}
%%%%%%%%%%%%%%%%%%%%%%%%%%%%%%%%%%%%%%
with
%%%%%%%%%%%%%%%%%%%%%%%%%%%%%%%%%%%%%%
\begin{align}
&\hat{H}_n(r,\partial_r)
=\xi_{n_i}(r,\partial_r) \sigma_0 \otimes I_{N \times N} + \hat{A}_n(r,\partial_r),
\\
%%%%%%%%%%%%%%%%%%%%%%%%%%%%
&-\hat{H}_{-n}(r,\partial_r)
=-\xi_{l_i}(r,\partial_r) \sigma_0 \otimes I_{N \times N} -\hat{A}_{-n}(r,\partial_r),
\\
%%%%%%%%%%%%%%%%%%%%%%%%%%%
&\xi_{n_i}(r,\partial_r)
=-\left( \partial^2_r+\frac{1}{r}\partial_r-\frac{n^2_i}{r^2} \right)-\mu,
\\
%%%%%%%%%%%%%%%%%%%%%%%%%%%
&\hat{\Delta}(r)
=
\Biggl(
\begin{array}{cc}
f(r)i \sigma_y & 0 \\
0 & sf(r)i \sigma_y
\end{array}
\Biggr),
\end{align}
%%%%%%%%%%%%%%%%%%%%%%%%%%%%%%%%%
and
%%%%%%%%%%%%%%%%%%%%%%%%%%%%%%%%%
\begin{align}
\hat{A}_n(r,\partial_r)
=
\left(
\begin{array}{cc}
-h & -\frac{\alpha}{k_{\rm F}a}\left( \partial_r+\frac{n+1}{r} \right)  \\
\frac{\alpha}{k_{\rm F}a}\left( \partial_r-\frac{n}{r} \right) & h   \\
t_\perp & 0\\
 0 & t_\perp
\end{array}
\right.& \nonumber \\
\left.
\begin{array}{cc}
 t_\perp & 0  \\
 0& t_\perp \\
 -h &\frac{\alpha}{k_{\rm F}a} \left( \partial_r+\frac{n+1}{r} \right)  \\
-\frac{\alpha}{k_{\rm F}a} \left( \partial_r -\frac{n}{r}\right)& h
\end{array}
\right),&
\end{align}
%\begin{array}{cccc}
%-h & -\frac{\alpha}{k_{\rm F}\xi_0}\left( \partial_r+\frac{n+1}{r} \right)& t_\perp & 0 \\
%\frac{\alpha}{k_{\rm F}\xi_0}\left( \partial_r-\frac{n}{r} \right) & h & 0& t_\perp \\
%t_\perp & 0&-h &\frac{\alpha}{k_{\rm F}\xi_0} \left( \partial_r+\frac{n+1}{r} \right)  \\
%0 & t_\perp& -\frac{\alpha}{k_{\rm F}\xi_0} \left( \partial_r -\frac{n}{r}\right)& h 
%\end{array}
%\right),
%\end{align}
%%%%%%%%%%%%%%%%%%%%%%%%%%%%%%%%%
%where we drop the terms that do not alter the quantum number of the orbital angular momentum in the kinetic energy term.
%where we have put $\mu_{\rm B}H/\Delta_0 \rightarrow h$, $\alpha/\Delta_0 \rightarrow \alpha$, $r/\xi_0 \rightarrow r$, $t_\perp/\Delta_0 \rightarrow t_\perp$ and $E/\Delta_0 \rightarrow E$.
where we have put $\mu/t \rightarrow \mu$, $\mu_{\rm B}H/t \rightarrow h$, $\alpha/t \rightarrow \alpha$, $r/a \rightarrow r$, $t_\perp/t \rightarrow t_\perp$, $E/t \rightarrow E$ and $k_{\rm F}a=1$.

%We take the cut-off radius as $r_{\rm c}=20\xi_0$ and discretize the BdG equation in the real space by the mesh size $0.125\xi_0$ (mesh number $N=160$).
%Then, we diagonalize the $2^3N \times 2^3 N$ BdG Hamiltonian with respect to each quantum number of the angular momentum $n$.
%For each $n$,
%we arrange the eigen energy $E_{jn}$ $(0<|j|<2^3N)$ in ascending order for later discussions.

We take the cut-off radius as $r_{\rm c}=250a$ and discretize the BdG equation in the real space by the mesh size $0.625a$ (mesh number $N=400$).
Then, we diagonalize the $2^3N \times 2^3 N$ BdG Hamiltonian with respect to each quantum number of the angular momentum $n$.
For each $n$,
we arrange the eigen energy $E_{jn}$ $(0<|j|<2^3N)$ in ascending order for later discussions.

%In the self-consistent calculation,
%the pair potential profile drops to zero even in the bulk if the maximum quantum number of the angular momentum $n_{\rm max}$ is not sufficiently large.
%So one can know the safe choice of $n_{\rm max}$.
%In our calculation using the test potential,
%we roughly estimate the angular momentum quasiclassically as $|{\vecbm L}_z|=(k_{\rm F}\xi_0)(r/\xi_0)$ and
%set $n_{\rm max}=60$ for $k_{\rm F}\xi_0=3$ and $r_{\rm_c}=20\xi_0$.

%The local density of states is calculated from
%%%%%%%%%%%%%%%%%%%%%%%%%%%%%%%%%
%\begin{equation}
%N({\vecbm r},E)=-\sum_{jn,\sigma m}\left[ |u^{\sigma m}_{jn}({\vecbm r})|^2 f^\prime(E-E_{jn})
%+|v^{\sigma m}_{jn}({\vecbm r})|^2 f^\prime(E+E_{jn}) \right]
%\end{equation}
%%%%%%%%%%%%%%%%%%%%%%%%%%%%%%%%%
%with the Fermi function $f(E)=[1+\exp\{(E-\mu)/T\}]^{-1}$.

%%%%%%%%%%%%%%%%%%%%%%%%%%%%%%%%%%%%%%%%%%%%%%%%%%%%%%%%%%%%%%%%%%%%%%%%%%%%%%%%%%%%%%%%%%%%%%%%%%%%%%%%%%%%%%%%%%%%%%%%%%%%%%%%%%%%%%
%%%%%%%%%%%%%%%%%%%%%%%%%%%%%%%%%%%%%%%%%%%%%%%%%%%%%%%%%%%%%%%%%%%%%%%%%%%%%%%%%%%%%%%%%%%%%%%%%%%%%%%%%%%%%%%%%%%%%%%%%%%%%%%%%%%%%%
\section{Results and discussions}
%%%%%%%%%%%%%%%%%%%%%%%%%%%%%%%%%
\begin{figure}[tb]
\begin{center}
\begin{tabular}{p{80mm}p{80mm}p{80mm}}
      \resizebox{80mm}{!}{\includegraphics{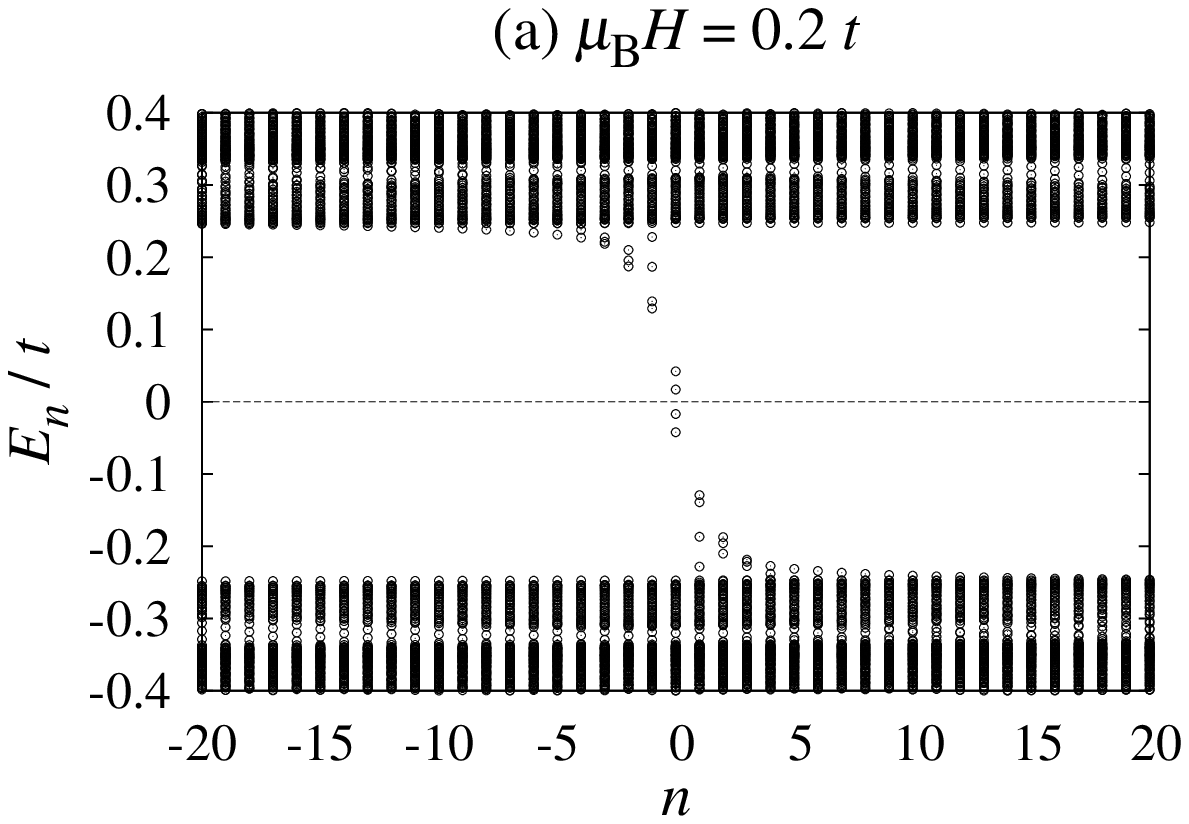}} \\
      \resizebox{80mm}{!}{\includegraphics{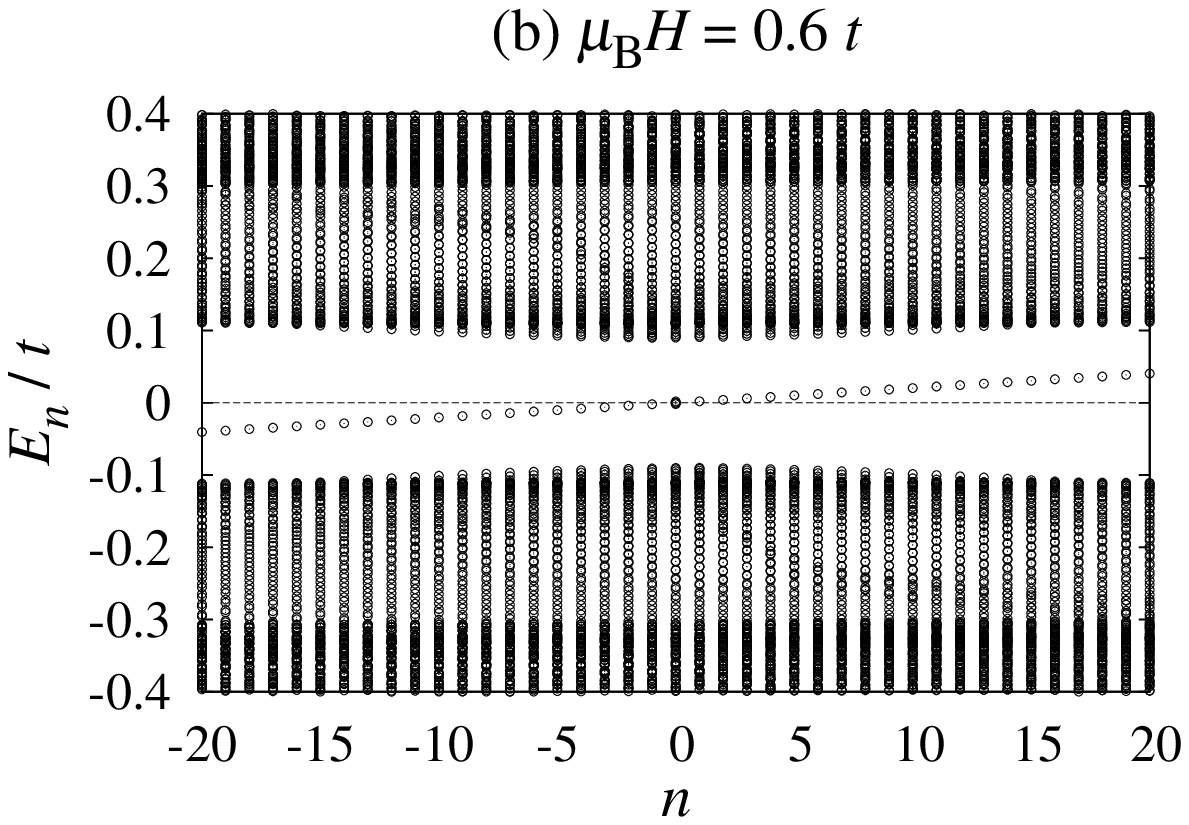}} \\
      \resizebox{80mm}{!}{\includegraphics{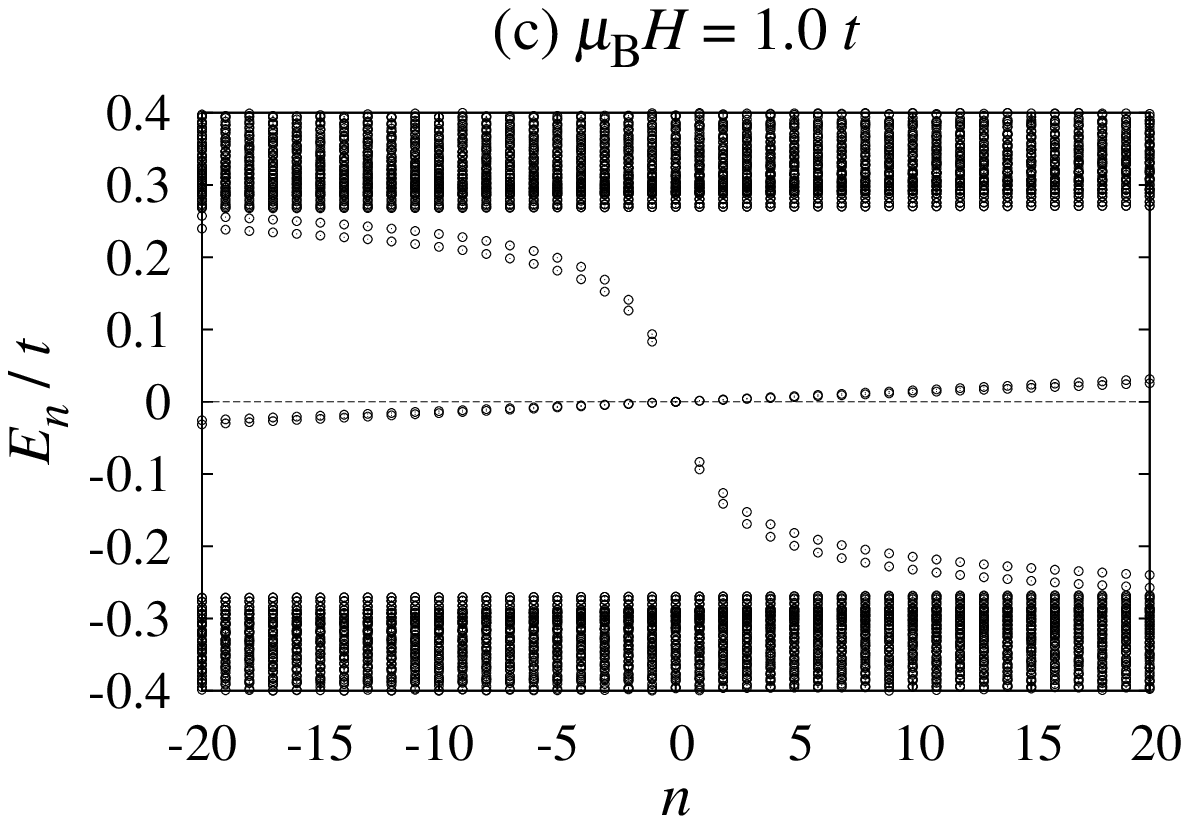}} \\
\end{tabular}
\caption{
\label{fig1}
Energy spectra of Bogoliubov quasiparticles in the PDW state for (a) $\mu_{\rm B}H/t=0.2$, (b) $0.6$ and (c) $1$.
The horizontal axis $n$ is the quantum number of the angular momentum.
We set $\alpha/t=1$, $t_\perp/t=0.1$, $\varDelta_0/t=0.35$ and $\mu/t=0.5$.
The dotted lines indicate the zero energy.
}
\end{center}
\end{figure}
%%%%%%%%%%%%%%%%%%%%%%%%%%%%%%%%%

%%%%%%%%%%%%%%%%%%%%%%%%%%%%%%%%%
\begin{figure}[tb]
\begin{center}
\begin{tabular}{p{80mm}}
      \resizebox{80mm}{!}{\includegraphics{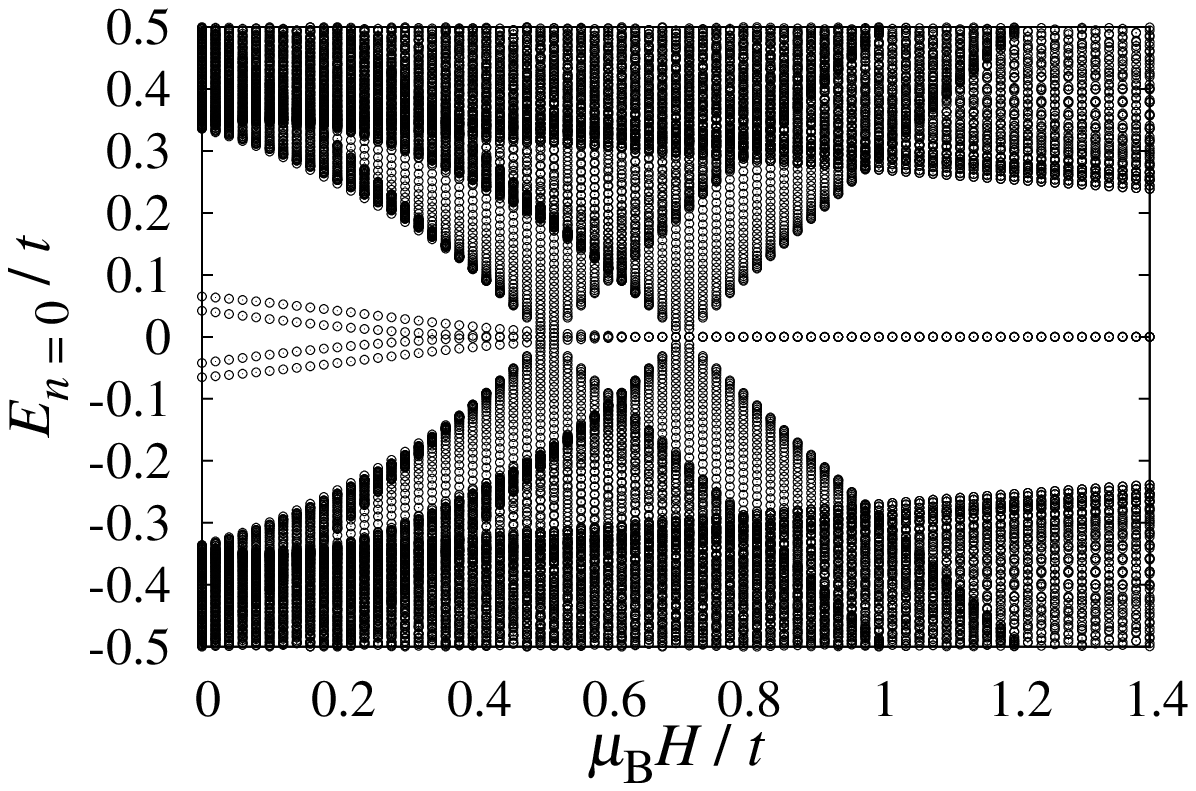}} \\
\end{tabular}
\caption{
\label{fig2}
Magnetic field dependence of the energy spectra of Bogoliubov quasiparticles with $n=0$ in the PDW state.
We set $\alpha/t=1$, $t_\perp/t=0.1$, $\varDelta_0/t=0.35$ and $\mu/t=0.5$.
}
\end{center}
\end{figure}
%%%%%%%%%%%%%%%%%%%%%%%%%%%%%%%%%

%%%%%%%%%%%%%%%%%%%%%%%%%%%%%%%%%
\begin{figure}[tb]
\begin{center}
\begin{tabular}{p{80mm}p{80mm}p{80mm}}
      \resizebox{80mm}{!}{\includegraphics{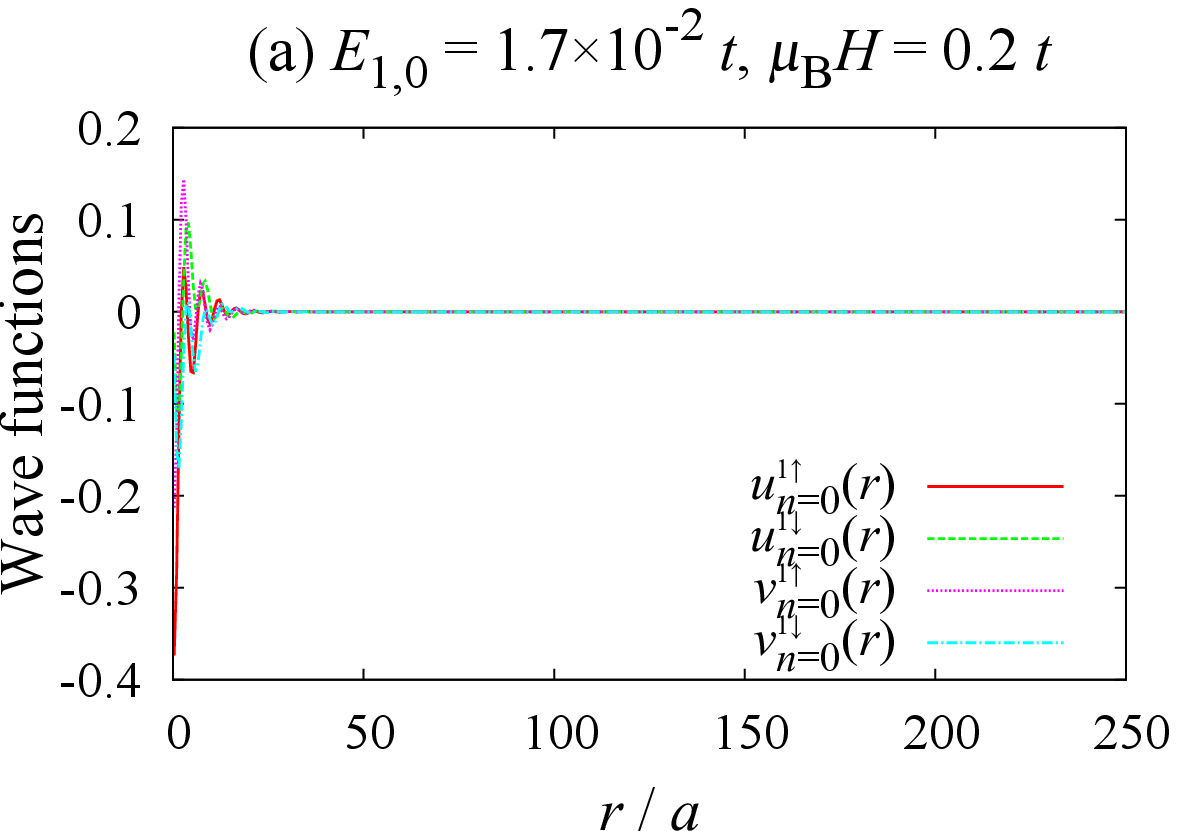}} \\
      \resizebox{80mm}{!}{\includegraphics{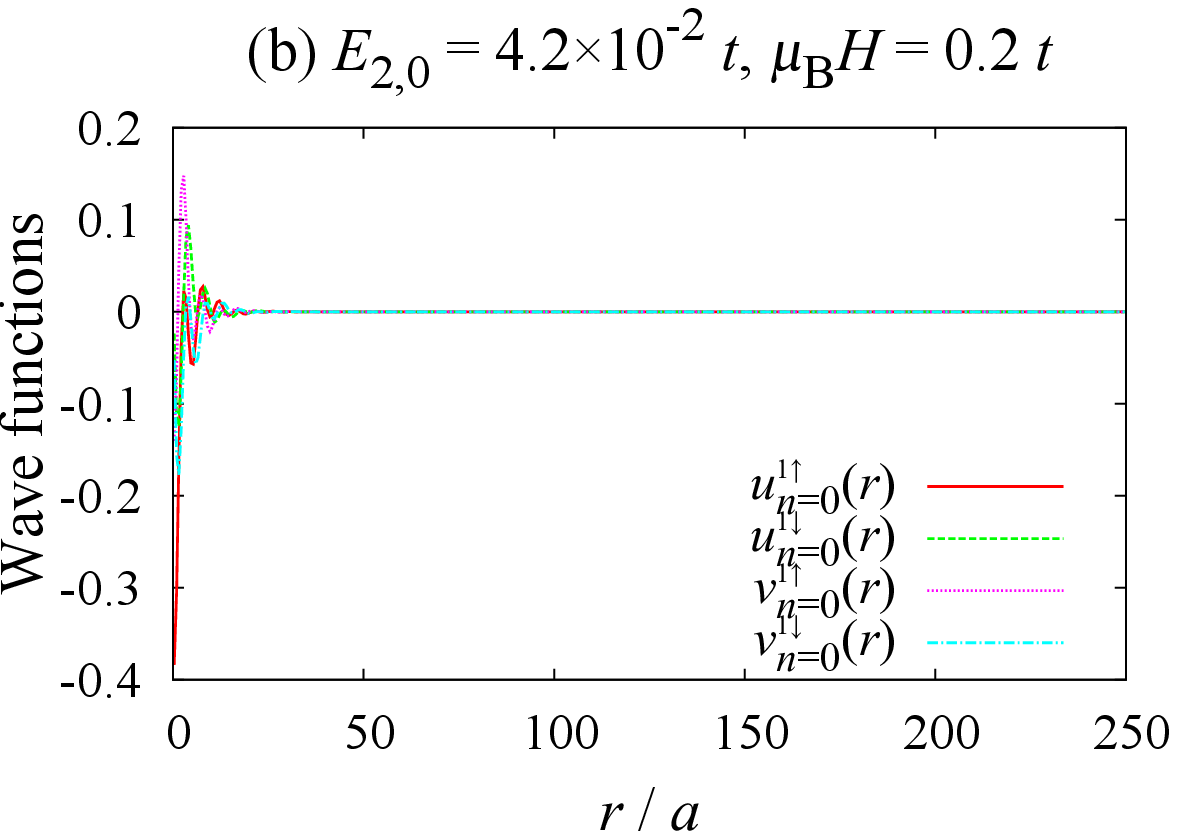}} \\
      \resizebox{80mm}{!}{\includegraphics{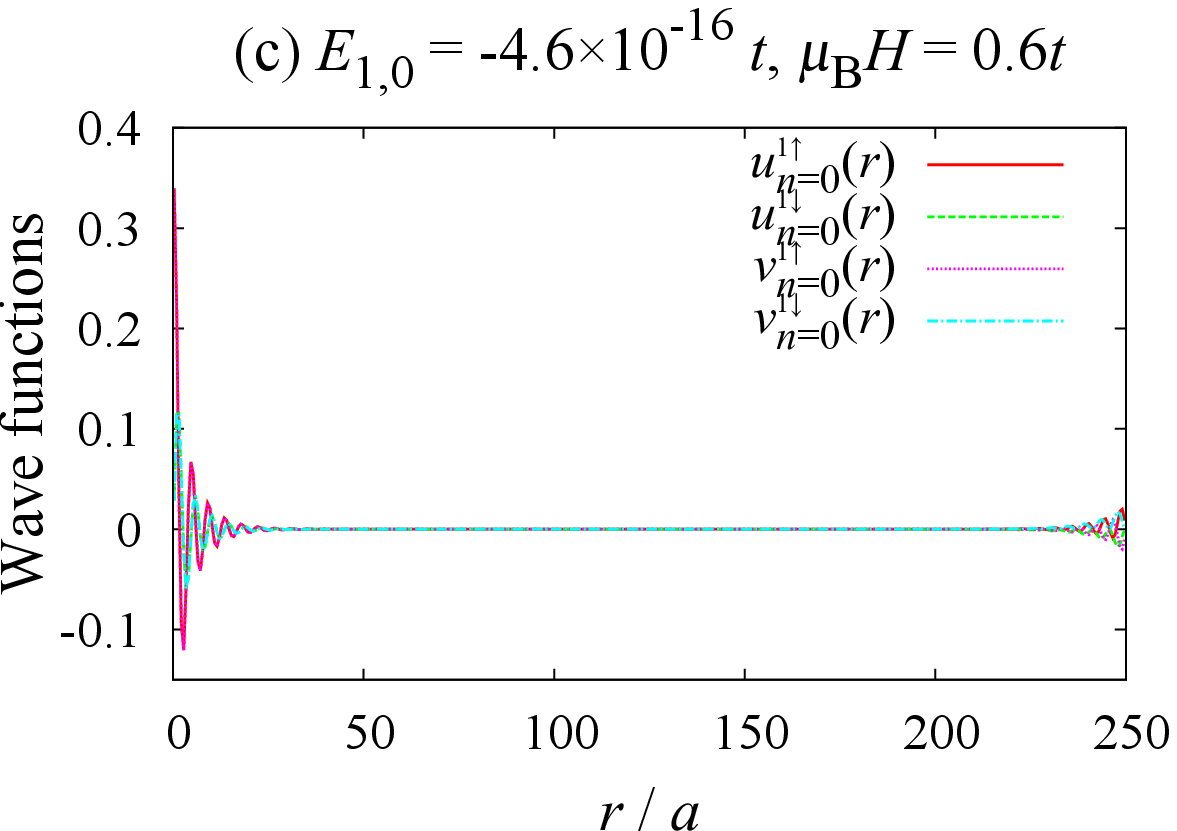}} 
    \end{tabular}
%    \caption{
%\label{fig3}
%Wave functions of Bogoliubov quasiparticles in the PDW state belonging to the lowest eigen energy $E_{1,0}$ and $E_{2,0}$ for $\mu_{\rm B}H/t=0.2$ [(a), (b)], $\mu_{\rm B}H/t=0.4$ [(c), (d)]
%and $\mu_{\rm B}H/t=1$ [(e), (f)].
%We set $\alpha/t=1$, $t_\perp/t=0.1$, $\varDelta_0/t=0.35$ and $\mu/t=0.5$.
%} 
\end{center}
\end{figure}
%%%%%%%%%%%%%%%%%%%%%%%%%%%%%%%%%

%%%%%%%%%%%%%%%%%%%%%%%%%%%%%%%%%
\begin{figure}[tb]
\begin{center}
\begin{tabular}{p{80mm}p{80mm}p{80mm}}
      \resizebox{80mm}{!}{\includegraphics{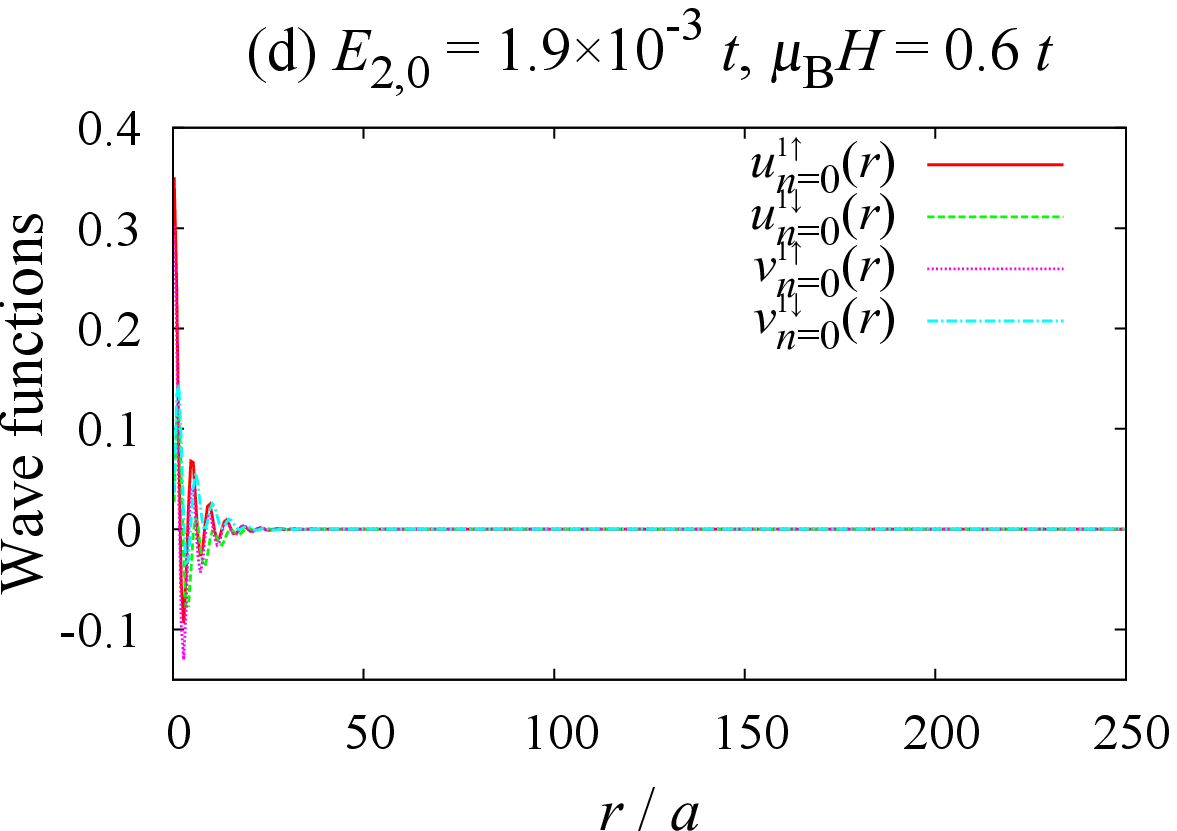}} \\
      \resizebox{80mm}{!}{\includegraphics{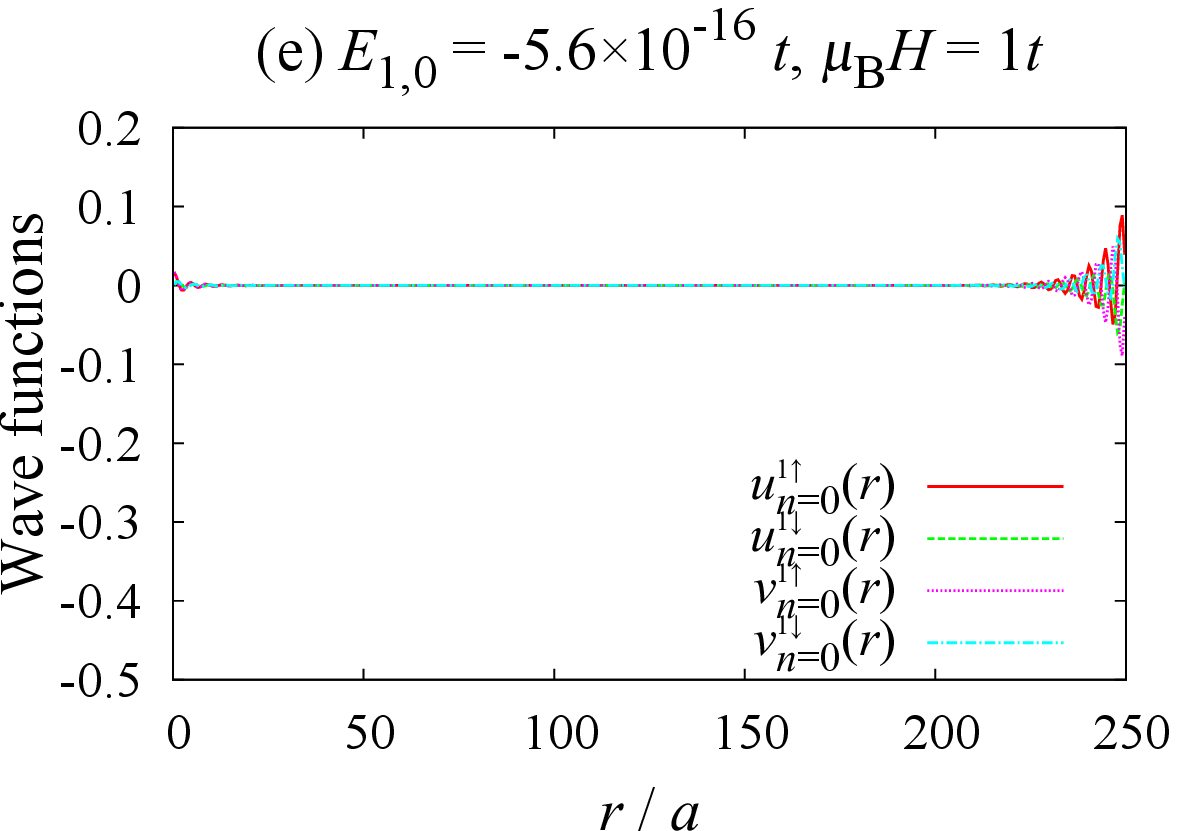}} \\
      \resizebox{80mm}{!}{\includegraphics{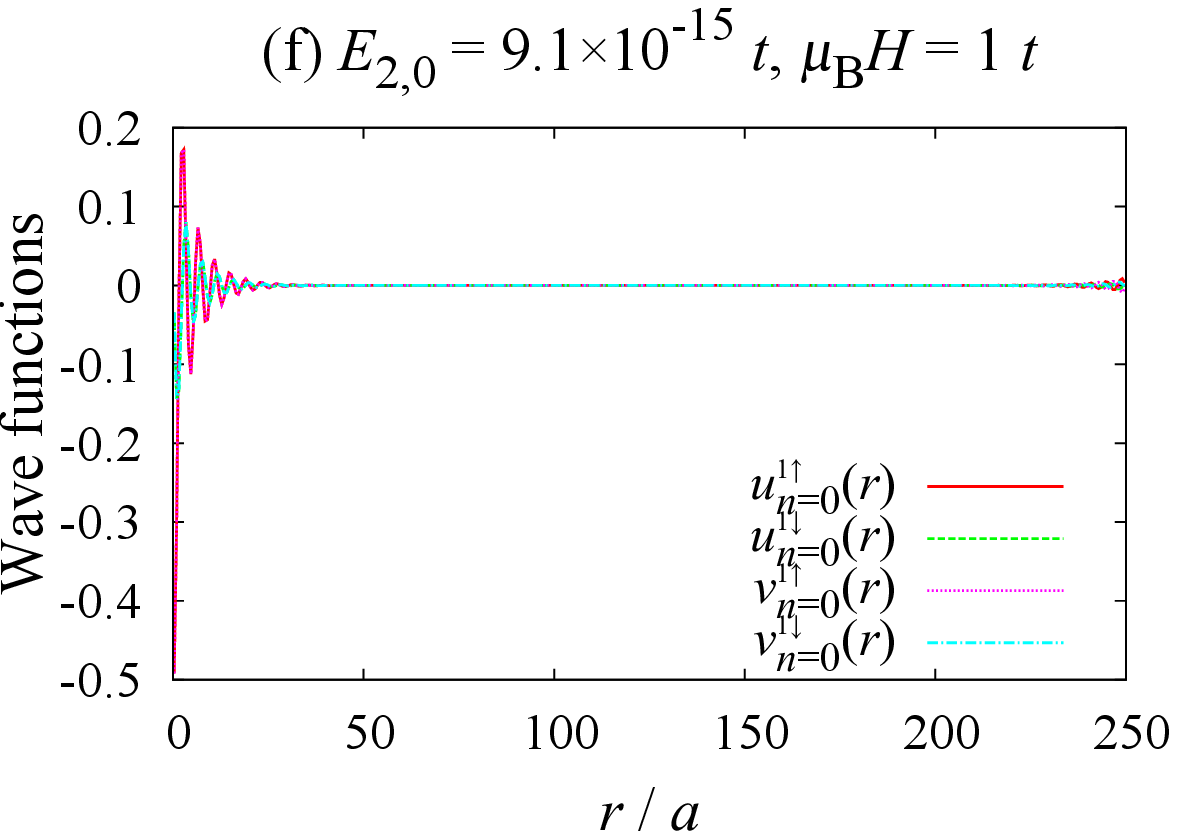}} 
    \end{tabular}
    \caption{
\label{fig3}
Wave functions of Bogoliubov quasiparticles in the PDW state belonging to the lowest eigen energy $E_{1,0}$ and $E_{2,0}$ for $\mu_{\rm B}H/t=0.2$ [(a), (b)], $\mu_{\rm B}H/t=0.4$ [(c), (d)]
and $\mu_{\rm B}H/t=1$ [(e), (f)].
We set $\alpha/t=1$, $t_\perp/t=0.1$, $\varDelta_0/t=0.35$ and $\mu/t=0.5$.
} 
\end{center}
\end{figure}
%%%%%%%%%%%%%%%%%%%%%%%%%%%%%%%%%

Using the mirror symmetry of the system,
the BdG Hamiltonian is block-diagonalized \cite{Ueno2013} and each subsector becomes the Hamiltonian of the noncentrosymmetric superconductor
with the Rashba ASOC under effective magnetic fields $\mu_{\rm B}H \pm t_\perp$ \cite{Yoshida-unpub2014}.
This system undergoes a topologically non-trivial phase under the sufficiently large effective magnetic field
$|\mu_{\rm B}H\pm t_\perp| > \sqrt{\mu^2+\Delta^2_0}$
%$\left[|(\mu_{\rm B}H \pm t_\perp)/\Delta_0|>\sqrt{(k_{\rm F}\xi_0/2)^2+1}\right]$
\cite{Yoshida-unpub2014,Sato2009,Tewari2009}.
%There are two critical magnetic field values, $h^+_{\rm c}\approx 0.8$ and $h^-_{\rm c}\approx2.8$ for $k_{\rm F}\xi_0=3$.
There are two critical magnetic field values, $h^+_{\rm c}\approx 0.5$ and $h^-_{\rm c}\approx0.7$
%for $k_{\rm F}\xi_0=3$.
for $\mu/t=0.5$, $\varDelta_0/t=0.35$ and $t_\perp/t=0.1$.

First, we show in Fig.~\ref{fig1} the energy spectra of Bogoliubov quasiparticles in the PDW state for $\alpha/t=1$, $t_\perp/t=0.1$, $\mu/t=0.5$ and $\varDelta_0/t=0.35$.
In the intermediate magnetic field [$h^+_{\rm c}<h~(=0.6)<h^-_{\rm c}$], as shown in Fig.~\ref{fig1}(b),
a single branch of the edge mode appears inside the gap.
This is because one mirror sector with the effective magnetic field $\mu_{\rm B}H+t_\perp$ enters the topologically non-trivial phase.
For the low field [e.g., $h~(=0.2)<h^+_{\rm c}$], no edge mode appears [See Fig.~\ref{fig1}(a)].
On the other hand, in the high magnetic field [$h^-_{\rm c}<h~(=1)$],
two branches of the edge mode appear inside the gap as displayed in Fig.~\ref{fig1}(c).
This indicates that both mirror sectors satisfy the condition of the topologically non-trivial phase.
We can also see the vortex bound states [Caroli-de Gennes-Matricon (CdGM) modes] within the superconducting gap in Figs.~\ref{fig1}(a) and \ref{fig1}(c).
In Fig.~\ref{fig1}(b), the CdGM modes are hidden in the continuum spectra except for $E_{1,0}$, since the superconducting gap becomes small in the vicinity of the two critical magnetic fields.
The negative slopes of CdGM modes reflect the direction of a magnetic field [$\varDelta({\vecbm r})=\varDelta_0(r) \exp(i\phi_{\rm r})$].
%We can also see the vortex bound states [Caroli-de Gennes-Matricon (CdGM) modes] within the superconducting gap in Figs.~\ref{fig1}(a) and \ref{fig1}(b).
%Their negative slopes reflect the direction of a magnetic field [$\varDelta({\vecbm r})=\varDelta_0(r) \exp(i\phi_{\rm r})$].

Fig.~\ref{fig2} shows the magnetic field dependence of the energy spectra for $n=0$.
At the critical field values [$h^+_{\rm c}\approx 0.5$ and $h^-_{\rm c}\approx0.7$],
the superconducting gap closes.
In the low field $h<h^+_{\rm c}$,
there are two finite eigen energies inside the gap in the positive energy side.
The lowest one comes from the mirror sector under the effective magnetic field $\mu_{\rm B}H+t_\perp$
and the other one comes from that with $\mu_{\rm B}H-t_\perp$.
Then, at $h^+_{\rm c}\approx 0.5$,
the mirror sector under the effective magnetic field $\mu_{\rm B}H+t_\perp$ enters the topological phase.
Subsequently,
at $h^-_{\rm c}\approx0.7$,
the mirror sector under the effective magnetic field $\mu_{\rm B}H-t_\perp$ undergoes the phase transition into the topological phase.
In $h^+_{\rm c}<h<h^-_{\rm c}$, the mirror sector with $\mu_{\rm B}H+t_\perp$ has the strictly zero eigen energy
and in $h>h^-_{\rm c}$, the both mirror sectors have the strictly zero eigen energy.

Next,
we investigate the wave functions belonging to the two lowest eigen energy with $n=0$.
Figs.~\ref{fig3}(a) and \ref{fig3}(b) show the wave functions with their eigen energy $E_{1,0} \approx 1.7\times 10^{-2}t$ and $E_{2,0} \approx 4.2 \times 10^{-2}t$, respectively for $\mu_{\rm B}H/t=0.2$.
We show the wave functions of quasiparticles in the layer 1 only.
The quasiparticle wave functions are localized around the vortex and have no amplitude at the edge.
This is consistent with Fig.~\ref{fig1}(a), in which only the vortex bound states appear inside the gap.
Then, we consider the intermediate magnetic field regime [$h^+_{\rm c}<h~(=0.6)<h^-_{\rm c}$].
In Fig.~\ref{fig3}(d),
the wave functions with the energy $E_{2,0}\approx 1.9\times 10^{-3}t$ are localized at the vortex core and have no amplitude at the edge for $\mu_{\rm B}H/t=0.6$.
We can consider that these wave functions are the eigen states of one mirror sector with the effective magnetic field $\mu_{\rm B}H-t_\perp$,
that is, the eigen states of the mirror sector in the topologically trivial phase.
%Note that the eigen energy $E_{1,0}$ accidentally becomes nearly zero ($\approx 7.1\times 10^{-16}\Delta_0$),
%since the zero energy eigen states of the sector with the effective field $(\mu_{\rm B}H-t_\perp)/\Delta_0=0.5$ are not topologically protected for $h=1.5$.
%Reflecting the odd parity superconductivity in the PDW state \cite{Yoshida2012, Yoshida2014},
%the exactly zero energy eigen energy $E_{1,0} \approx 7.1\times 10^{-16}\Delta_0$
%We obtain the effective gap function 
%$\Delta^{\rm eff}({\vecbm r},\tilde{{\vecbm k}})={\vecbm d}^{\rm eff}_k({\vecbm r}) \cdot {\vecbm \sigma}i \sigma_y$
%with ${\vecbm d}^{\rm eff}_k({\vecbm r})=\left[ -\alpha/\sqrt{t^2_\perp+\alpha^2|{\vecbm g}(\tilde{{\vecbm k}})|^2} \right] \Delta({\vecbm r}) {\vecbm g}(\tilde{{\vecbm k}})$,
%following the same procedure described in Ref.~\cite{Nagai2014}.
%We will discuss the detail elsewhere.
On the other hand, as show in Fig.~\ref{fig3}(c),
the wave functions with the energy $E_{1,0}\approx-4.6 \times 10^{-16}t$ (i.e., zero energy) are localized both at the vortex core and at the edge.
These eigen wave functions correspond to those of one mirror sector with the effective field $\mu_{\rm B}H+t_\perp \approx 0.7t$,
which is in the topologically non-trivial phase.
The edge bound states appear also in the energy spectra in Fig.~\ref{fig1}(b).
Next,
we consider the high magnetic field regime [$h^-_{\rm c}<h~(=1.0)$].
In this situation,
both the mirror sectors are in the topologically non-trivial phase,
and so the wave functions with the lowest eigen energies $E_{1,0}\approx-5.6 \times 10^{-16}t$ and $E_{2,0}\approx 9.1 \times 10^{-15}t$ (i.e., zero energies)
have the amplitude both at the vortex core and at the edge [see Figs.~\ref{fig3}(e) and (f)].

%%%%%%At last,
%%%%%%we comment on the point that the lowest eigen energy has the small finite value in Figs.~\ref{fig2}(b)-\ref{fig2}(d).
%The minigap in the energy level of the vortex bound states in an $s$-wave superconductor is estimated as $\Delta^2_0/\varepsilon_{\rm F} \sim \Delta_0/(k_{\rm F}\xi_0)$ \cite{Caroli1964}.
%We obtain $\Delta^2_0/\varepsilon_{\rm F} \approx 0.33 \Delta_0$ for $k_{\rm F}\xi_0=3$ in the present system. 
%This value is much larger than the value of the two lowest eigen energies $E_{1,0}$ and $E_{2,0}$ in Figs.~\ref{fig2}(b)-\ref{fig2}(d). 
%Then, 
%we attribute the above mentined point to the finite size of the system.
%%%%%%We attribute this fact to the finite size of the system.
%%%%%%It is thought that the lowest eigen energy goes to zero in the thermodynamic limit.
%%%%%%In other words,
%%%%%%the wave functions of the CdGM modes and the edge mode are connected with each other in Figs.~\ref{fig2}(b)-\ref{fig2}(d).
%%%%%%Owing to the hybridization of the quasiparticle wave function between the CdGM mode and the edge mode,
%%%%%%the zero energy eigen energy shift to the significant finite value in Figs. \ref{fig2}(b)-\ref{fig2}(d).

%%%%%%%%%%%%%%%%%%%%%%%%%%%%%%%%%%%%%%%%%%%%%%%%%%%%%%%%%%%%%%%%%%%%%%%%%%%%%%%%%%%%%%%%%%%%%%%%%%%%%%%%%%%%%%%%%%%%%%%%%%%%%%%%%%%%%%
%%%%%%%%%%%%%%%%%%%%%%%%%%%%%%%%%%%%%%%%%%%%%%%%%%%%%%%%%%%%%%%%%%%%%%%%%%%%%%%%%%%%%%%%%%%%%%%%%%%%%%%%%%%%%%%%%%%%%%%%%%%%%%%%%%%%%%
\section{Conclusion}
We have formulated the bilayer Rashba superconductors in the presence of a vortex by means of Bogoliubov-de Gennes theory.
We have investigated the energy spectra and the wave functions of bound Bogoliubov quasiparticles at a vortex and an edge in the pair-density wave state.
Our calculations of the energy spectra confirm that the branches of the edge mode appear inside the gap if satisfying the condition of the topologically non-trivial phase.
We can confirm the appearance of the zero energy vortex and edge quasiparticle states also from the spatial profiles of the zero energy eigen wave functions
under the magnetic field above the critical values.

%%%%%%%%%%%%%%%%%%%%%%%%%%%%%%%%%%%%%%%%%%%%%%%%%%%%%%%%%%%%%%%%%%%%%%%%%%%%%%%%%%%%%%%%%%%%%%%%%%%%%%%%%%%%%%%%%%%%%%%%%%%%%%%%%%%%%%
%%%%%%%%%%%%%%%%%%%%%%%%%%%%%%%%%%%%%%%%%%%%%%%%%%%%%%%%%%%%%%%%%%%%%%%%%%%%%%%%%%%%%%%%%%%%%%%%%%%%%%%%%%%%%%%%%%%%%%%%%%%%%%%%%%%%%%
\section*{Acknowledgments}
The authors thank T. Kawakami and Y. Masaki for helpful discussions.
This study has been partially supported by JSPS KAKENHI Grant Number 26400367.
%\section*{References}
\bibliographystyle{elsarticle-num}
\bibliography{<your-bib-database>}



% \end{thebibliography}

\end{document}